\documentclass[conference]{IEEEtran}
\IEEEoverridecommandlockouts
\usepackage{threeparttable}
\usepackage{algorithm}
\usepackage{algorithmicx}  
\usepackage{algpseudocode}
\usepackage{float}
\usepackage{cite}
\usepackage{amsmath,amssymb,amsfonts}
\usepackage{inputenc}
\usepackage{graphicx}
\usepackage{subfigure}
\usepackage{textcomp}
\usepackage{xcolor}
\usepackage{url}
\usepackage{multirow}
\usepackage{booktabs}%this package is responsible for drawing table
\usepackage{xcolor}
\def\BibTeX{{\rm B\kern-.05em{\sc i\kern-.025em b}\kern-.08em
    T\kern-.1667em\lower.7ex\hbox{E}\kern-.125emX}}
\begin{document}

\title{HTDet: A Clustering Method using Information Entropy for Hardware Trojan Detection}
\author{
    \IEEEauthorblockN{
    Renjie Lu\IEEEauthorrefmark{1}, Haihua Shen\IEEEauthorrefmark{1}, Feng Zhang\IEEEauthorrefmark{2}, Huawei Li\IEEEauthorrefmark{3}, Wei Zhao\IEEEauthorrefmark{1}, and Xiaowei Li\IEEEauthorrefmark{3}
    }
    \IEEEauthorrefmark{1}University of Chinese Academy of Sciences\\
    \IEEEauthorrefmark{2}Institute of Microelectronics, Chinese Academy of Sciences\\
    \IEEEauthorrefmark{3}Institute of Computing Technology, Chinese Academy of Sciences
    \IEEEauthorblockA{Beijing, China}
    \thanks{Corresponding author: shenhh@ucas.ac.cn; lihuawei@ict.ac.cn}

 }

\maketitle

\begin{abstract}
 Hardware Trojans (HTs) have drawn more and
more attention in both academia and industry because of its significant potential threat. In this paper, we proposed a novel HT detection method using information entropy based clustering, named HTDet. The key insight of HTDet is that the Trojan usually be inserted in the regions with low controllability and low observability in order to maintain high concealment, which will result in that Trojan logics appear extremely low transitions during the simulation. This means that the logical regions with the low transitions will provide us with much more abundant and more important information for HT detection. Therefore, HTDet applies information theory technology and a typical density-based clustering algorithm called Density-Based Spatial Clustering of Applications with Noise (DBSCAN) to detect all suspicious Trojan logics in circuit under detection (CUD). DBSCAN is an unsupervised learning algorithm, which can improve the applicability of HTDet. Besides, we develop a heuristic test patterns generation method using mutual information to increase the transitions of suspicious Trojan logics. Experimental evaluation with benchmarks demenstrates the effectiveness of HTDet.
\end{abstract}


\section{Introduction}
Recently, the globalization of modern integrated circuit (IC) industry has raised more and more hardware security challenges. For example, intellectual property (IP) cores and EDA tools provided by the third-party are widely used in IC design to reduce development cost and to shorten the marketing cycle \cite{r1}. As third-party IP cores are designed by outsourced vendors, an adversary can easily implement some malicious logics into IP cores, referred to as Hardware Trojans (HTs).

HTs are lightweight structures in large-scale IC designs, which commonly contain two components called Trojan trigger and Trojan payload \cite{r2}. Trojan trigger is responsible for monitoring signals to determine whether the trigger signal has arrived. If Trojan trigger is not activated, HTs stay dormant and do not have effect on the original circuit. If Trojan trigger is activated, Trojan payload will perform specific malicious operations such as to change functionality, to degrade performance and to reveal secret information \cite{r3}. Since most of HTs usually have extremely rare trigger conditions, it is very challenging to detect suspicious Trojan logics in circuit under detection (CUD).

The existing HTs detection techniques can be roughly classified into five major groups: reverse engineering \cite{r4,r5,r6}, side channel analysis \cite{r7,r8,r9,r10,r11,r12,r13}, static structure analysis \cite{r15,r16,r17,r18,r19,r20}, statistical feature analysis \cite{r21,r22,r23,r24,r25}, and functional testing \cite{r26,r27,r28,r29}. In reverse engineering, a fabricated chip is completely dissected layer-by-layer in order to reconstruct the IC design to detect malicious modifications. Reverse engineering approaches consume prohibitively high cost, and it is impossible to carry out reverse engineering for each chip under test. In side-channel analysis, the impacts of HTs on circuit delay, transient current, leakage power and so on can be used to detect whether there are the HTs in CUD. Side-channel analysis approaches can detect HTs inserted in the post-fabrication stage. However, side-channel analysis usually requires a ``Golden Circuit" for impact comparison and it also is susceptible to process variations or environmental noise, which can result in lots of false positives. Like software virus detection technique, static structure analysis methods perform HT detection by analyzing circuit structure characteristics. Though static structure analysis is an effective HT detection approach, it can only detect known types of HTs. There are intrinsic differences between Trojan logics and normal circuit, so statistical feature analysis approaches can be used to detect potentail HTs in CUD.
Functional testing approaches try to generate test vectors to activate potential HTs and propagate HTs’ effects to the primary outputs. Though functional testing is independent with process variations and environmental noise, functional testing usually consume significant amount of time due to the high concealment of HTs.

The key insight of our approach is that Trojans usually be inserted in the regions with low controllability and low observability in order to maintain high concealment, which will result in that Trojan logics appear extremely low transitions during the simulation. In the field of information theory, if the event is improbable, it will provide much more information when the event happens. That is, the logical regions with the low transitions will provide us with much more abundant and more important information for Trojan detection. 
In this paper, we propose a novel HT detection method using information entropy based clustering, named HTDet. Firstly, the digital stimuli is generated for the CUD. Then the information entropy of signal sequence of each wire is calculated, and a typical density-based clustering algorithm called Density-Based Spatial Clustering of Applications with Noise (DBSCAN) is applied to obtain all suspicious Trojan logics. Further, a heuristic test patterns generation method using mutual information is developed to increase the transitions of these suspicious Trojan logics. In summary, this paper has the following contributions:
\begin{itemize}
\item To the best of our knowledge, this is the first attempt to use information entropy technology to detect HTs in hardware design, and HTDet can achieve good experimental results.
\item Unsupervised learning algorithm, DBSCAN, is used for Trojan detection, which means that HTDet does not require ``Golden Circuit". Further, HTDet does not require that the Trojan logic is pushed the triggering state. As long as the transitions of logical regions are extremely low, HTDet can detect them based on \textit{density-reachable} relationship.
\item We develop a heuristic test patterns generation method using mutual information technology to increase the transitions of suspicious Trojan logics.
\item We carry out lots of evalutaion work on TrustHub benchmarks \cite{r34}, which shows that the proposed technique can detect suspicious Trojan logics with negligible false positives.
\end{itemize}

The rest of this paper is organized as follows. Section 2 and Section 3 introduces the theoretical basis and the threat model, respectively. We present proposed HT detection method in detail in Section 4. Section 5 presents test patterns generation method for suspicious Trojan logics in detail. Experimental analysis is presented in Section 6. Section 7 briefly summarize the related work. Finally, we conclude this paper and in Section 8.

\section{Theoretical Basis}
In this paper, we perform the HT detection using information theory technology \cite{r30}. In this section, we give the theoretical basis of the proposed approach.

\subsection{Information Entropy}
Information Entropy is also known as the self-information, which is the average rate at which information is produced by a source of data. Entropy is a measure of uncertainty about random variable.

Let X be a discrete random variable, and its probability distribution is consistent with $p(x) = P(X=x)$, where $x \in X$. Hence, the entropy $H(X)$ of X can be explicitly written as
\begin{equation}
    H(X) = - \sum_{x \in X}p(x)\log_b p(x)
\end{equation}
, where b is the base of the logarithm used. In this paper, b is equal to the mathematical constant e. In the case of $p(x) = 0$, the value of $0\log_b 0$ is taken to be 0, which is consistent with the limit.
\begin{equation}
    \lim_{p(x) \to 0^{+}} p(x)\log_b p(x) = 0
\end{equation}

\subsection{Joint Entropy}
In information theory, joint entropy is a measure of the uncertainty associated with a set of variables. In this paper, we focus on the joint entropy of two random variables.

Similarly, let X and Y be two discrete random variables, and their probability distribution is $p(x,y)$, where $x \in X$ and $y \in Y$. Hence, the joint entropy $H(X, Y)$ of X， Y can be presented as
\begin{equation}
    H(X,Y) = - \sum_{x \in X}\sum_{y \in Y}p(x,y)\log_b p(x,y).
\end{equation}

\subsection{Conditional Entropy}
In information theory, the conditional entropy quantifies the amount of information needed to describe the outcome of a random variable Y given the value of another random variable X is known.

The entropy $H(Y|X)$ of Y conditioned on X can be defined as following formula.
\begin{equation}
\begin{aligned}
H(Y|X) = \sum_{x \in X}p(x)H(Y|X=x)\\
=\sum_{x \in X}p(x)\left[ -\sum_{y \in Y}p(y|x)\log_b p(y|x) \right]\\
= - \sum_{x \in X}\sum_{y \in Y}p(x,y)\log_b p(y|x)
\end{aligned}
\end{equation}

It's worth noting that $H(X)$, $H(X, Y)$ and $H(Y | X)$ can conform the chain rule. That is,
\begin{equation}
\begin{aligned}
  H(X,Y) = - \sum_{x \in X}\sum_{y \in Y}p(x,y)\log_b p(x,y)\\
  = - \sum_{x \in X}\sum_{y \in Y}p(x,y)\log_b \left[ p(x)p(y|x) \right]\\
  = - \sum_{x \in X}\sum_{y \in Y}p(x,y)\left[ \log_b p(x)+\log_b p(y|x) \right]\\
  = - \sum_{x \in X}p(x)\log_b p(x) - \sum_{x \in X}\sum_{y \in Y}p(x,y)\log_b p(y|x)\\
  = H(X) + H(Y|X)
\end{aligned}
\end{equation}

\subsection{Mutual Information}
The mutual information of two variables is a measure of the mutual dependence between the two variables. More specifically, the mutual information quantifies the amount of information obtained about one random variable through observing the other random variable.

Let X, Y be two discrete random variables, and their joint probability distribution is $p(x,y)$. Hence, the mutual information $I(X; Y)$ between X and Y can be defined as
\begin{equation}
    I(X;Y) = \sum_{x \in X}\sum_{y \in Y}p(x,y)\log_b \frac{p(x,y)}{p(x)p(y)}.
\end{equation}

According to the correlation between probability distributions and the chain rule, $I(X;Y)$ Can also be expressed as
\begin{equation}
\begin{aligned}
    I(X;Y) = \sum_{x \in X}\sum_{y \in Y}p(x,y)\log_b \frac{p(x|y)}{p(x)}\\
    = \sum_{x \in X}\sum_{y \in Y}p(x,y)\left[ \log_b p(x|y) - \log_b p(x) \right]\\
    = H(X) - H(X|Y) \\
    = H(X)+H(Y)-H(Y,X)
\end{aligned}
\end{equation}

\section{Threat Model}
The threat model of proposed method is based on several assumptions.
\begin{itemize}
\item With the globalization of chip design, the adversaries can have more opportunities to insert HTs into a digital circuit design than before. It can be gate-level netlist or register transfer language (RTL).
\item Our threat model assumes that the hardware design that we are given is in the form of digital circuit design.
\item The goal of attack is to change functionality, destroy the IC, and/or leak secret information through logical attack, rather than through side-channels such as current, power or electromagnetic.
\end{itemize}

\section{HTDet Methodology}
In this section, first, we provide the feasibility analysis of proposed HT detection method. Then the technical details of HTDet is presented. The core problem is whether the information entropy technology and clustering algorithm can be used to detect suspicious Trojan logics in the circuit under detection (CUD).

\subsection{Feasibility Analysis}
The key insight of HTDet is that there is the significant difference between the Trojan logic and the rest of the circuit. More specifically, the HT usually be inserted in the logical regions with low controllability and low observability, which causes that Trojan logic has a very low transition probability. Moreover, in the field of information theory \cite{r30}, if an event is very probable, the little information was provided when it happens. Conversely, if the event is improbable, it will provide much more information when the event happens.

That is, the logical regions with the low transitions will provide us with more abundant and more important information for HT detection. However, if we directly apply the transition probability for Trojan detection, which will result in high false positives. Fox example, we consider that the signal wires (from $W_1$ to $W_{14}$) have the transition probabilities listed in Table 1.

\begin{table*}[htbp]
\caption{Signal wires and corresponding transition probabilities.}
\begin{center}
\begin{tabular}{|c|c|c|c|c||c|c|c|c|c||c|c|c|c|c|}
\hline
Wire &	$W_1$	& $W_2$ & $W_3$	& $W_4$ &	$W_5$	& $W_6$ & $W_7$ & $W_8$ &	$W_9$	& $W_{10}$ & $W_{11}$ & $W_{12}$ & $W_{13}$	& $W_{14}$ \\
\hline
Transition Probability & $\frac{1}{1000}$ & $\frac{1}{800}$ & $\frac{1}{500}$ & $\frac{1}{200}$ & $\frac{1}{100}$ & $\frac{1}{80}$ & $\frac{1}{50}$ & $\frac{1}{20}$ & $\frac{1}{10}$ & $\frac{1}{8}$ & $\frac{1}{5}$ & $\frac{3}{10}$ & $\frac{1}{2}$ & $\frac{6}{10}$ \\
\hline
\end{tabular}
\end{center}
\end{table*}

Due to the \textit{density-reachable} relationship between low transition probability and high transition probability, signal wires from $W_1$ to $W_{10}$ can be reported as suspicious Trojan logic as shown in Figure 1 (blue line). While the use of information entropy can significantly reduce false positives. As shown in Figure 1 (orange line), signal wires from $W_1$ to $W_7$ can be reported as suspicious Trojan logic.
\begin{figure}[htbp]
\centerline{\includegraphics[width=7cm,height=5cm]{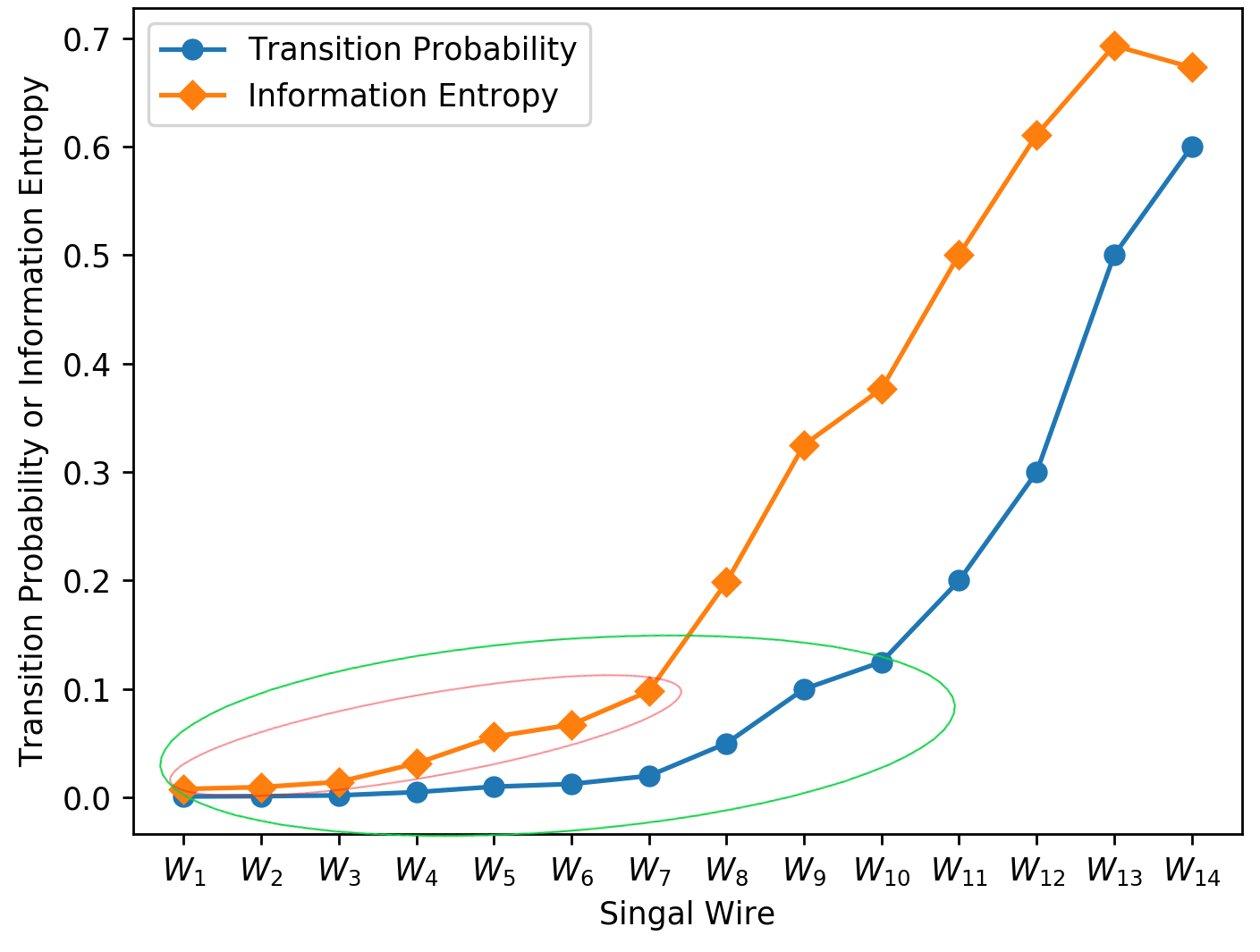}}
\caption{HT detection comparison between transition probability and information entropy.}
\label{fig}
\end{figure}

 This is because information entropy can gap the connectivity between low transition probability and high transition probability, and it is more sensitive to low transition probability as shown in Figure 2. It can be seen that the \textit{density-reachable} relationship between signal wires (from $W_1$ to $W_7$) is much closer than the \textit{density-reachable} relationship between low transition probability and high transition probability.
\begin{figure}[htbp]
\centerline{\includegraphics[width=7cm,height=5cm]{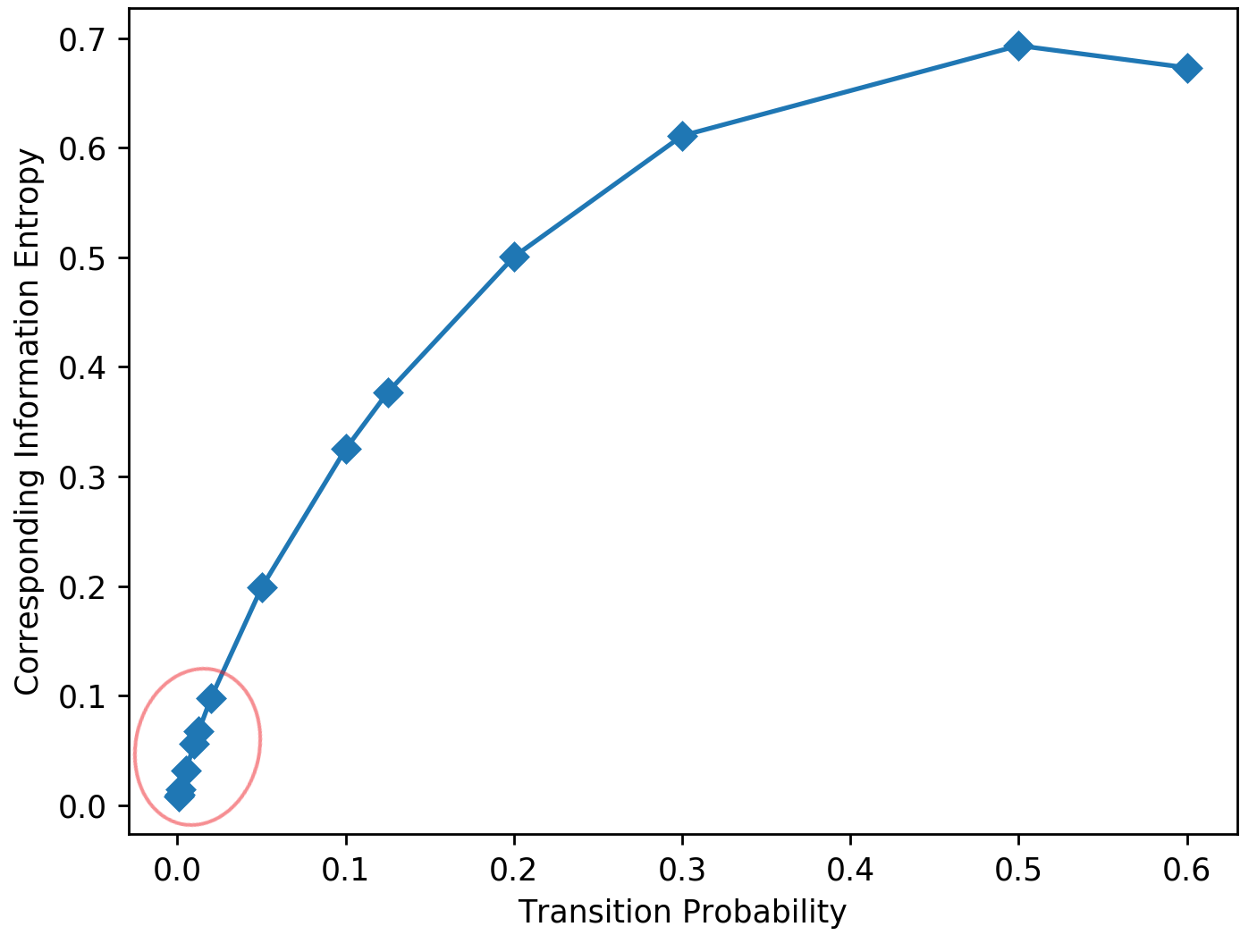}}
\caption{Distribution of information entropy for probability listed in Table 1.}
\label{fig}
\end{figure}

It has been proven that the information entropy takes the maximum value when $p(transition)$ is equal to $p(non-transition)$. In other words, when $p(transition)$ = $p(non-transition)$ = 0.5, the corresponding information entropy can take the maximum value. The transition probability-information entropy curve is as shown in Figure 3 according to formula (1). Because the information entropy has the symmetry, the minimum value can be taken when $p(transition)$ = 0 or $p(transition)$ = 1. Therefore, we should exclude the noise data that has very low information entropy because of much large $p(transition)$.
%since $p(transition)$ is closely equal to 1.
\begin{figure}[htbp]
\centerline{\includegraphics[width=7cm,height=5cm]{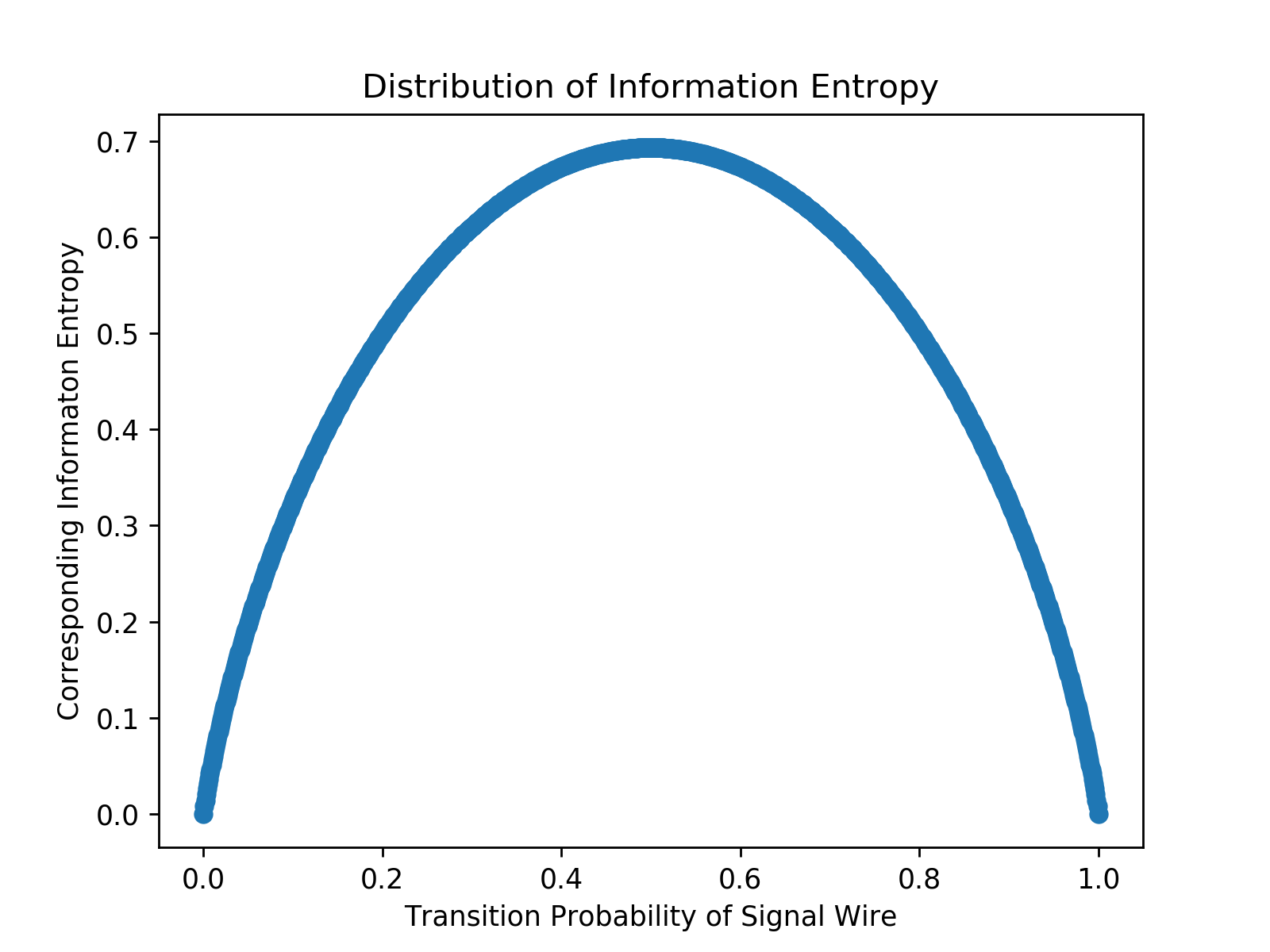}}
\caption{The transition probability-information entropy curve}
\label{fig}
\end{figure}

Besides, the mutual information technology can measure the correlations between primary inputs and internal signal wires, which is beneficial to test patterns generation. Therefore, we first propose applying the information theory technology in the field of HT detection.

\subsection{The Application of Information Entropy}
In order to apply the information entropy technology for HT detection, we first use functional testing to generate digital stimuli for the CUD. We believe that the set of test patterns developed during design verification can satisfy this step. The goal of this step is to perform functional tests for the CUD with high coverage as much as possible. After the functional tests, we can obtain the original waveform of each signal wire in the CUD, which contain only binary values (0 or 1). Our goal is to use the information entropy to evaluate the controllability and observability of each logical region such that we can effectively distinguish Trojan logic from the rest of circuit.

However, we can not use the original waveform for HT detection directly. For example, the transition of signal only occurs once in $OW_1$, while $OW_2$ have five transitions of signal as shown in Figure 4(a). Because the HT usually is inserted in the logical regions with a low controllability and low observability, which cause that the Trojan logic has a very low transition probability. Hence, the logical region of $OW_1$ should be more likely to be Trojan logic than $OW_2$. However, the information entropy of both $OW_1$ and $OW_2$ are 0.6931 according to formula (1) because the probability of 0 (0.5) and 1 (0.5) in $OW_1$ is the same as in $OW_2$.

\begin{figure}[htbp]
\centering
\subfigure[Original waveform $OW_1$ and $OW_2$.]{
\includegraphics[width=7cm,height=2cm]{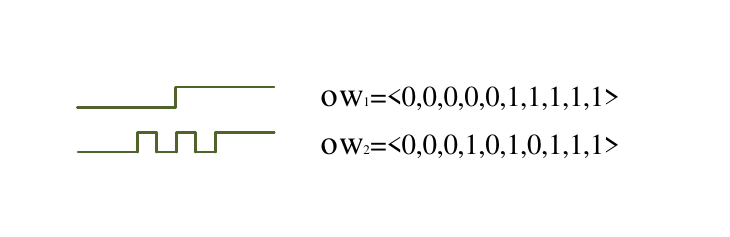}
}
\subfigure[Encoded waveform $Encode_1$ and $Encode_2$.]{
\includegraphics[width=7cm,height=2cm]{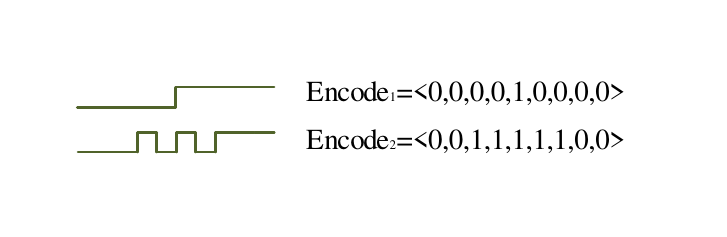}
}
\caption{Comparison between original waveform and encoded waveform.}
\end{figure}

Therefore, we should focus on the distribution of signal transitions rather than the distribution of 0 and 1 such that we can use the information entropy to evaluate the controllability and observability of each logical region. To this end, we encode the original waveform according to the following rules. We assume that the original waveform OW = $<s_1, s_2, ..., s_n, s_{n+1}>$. For each signal pair $<s_i, s_{i+1}>$, i = 1, 2, ..., n, if $<s_i, s_{i+1}>$ = $<0, 0>$, we encode $s_i$ as 0; if $<s_i, s_{i+1}>$ = $<0, 1>$, we encode $s_i$ as 1; if $<s_i, s_{i+1}>$ = $<1, 0>$, we encode $s_i$ as 1; if $<s_i, s_{i+1}>$ = $<1, 1>$, we encode $s_i$ as 0. The encoded waveform corresponding to the original waveform of $OW_1$ and $OW_2$ are shown in the Figure 4(b). Then, we use formula (1) to calculate the information entropy of each encoded waveform. The information entropy of $Encode_1$ (corresponding to $OW_1$) is approximately equal to 0.3488, and the information entropy of $Encode_2$ (corresponding to $OW_2$) is approximately equal to 0.6870, which is more in line with the results that we expect.

We apply the information entropy to distinguish differences between Trojan logic and the normal circuit. Lots of experiments demonstrate that the information entropy of each wire is almost consistent with the controllability measure \cite{r32} of this signal wire in the CUD. As shown in Figure 5, we can obtain information entropy of each wire in the given circuit after functional testing ($10^6$ cycles). It can be seen that the information entropy at the output of the AND gate is 0.13820, the information entropy at the input (top) of the AND gate is 0.22966 and the information entropy at the input ( bottom) of the AND gate is 0.66271 due to different circuit structures.

\begin{figure}[htbp]
\centerline{\includegraphics[width=8.5cm,height=4.8cm]{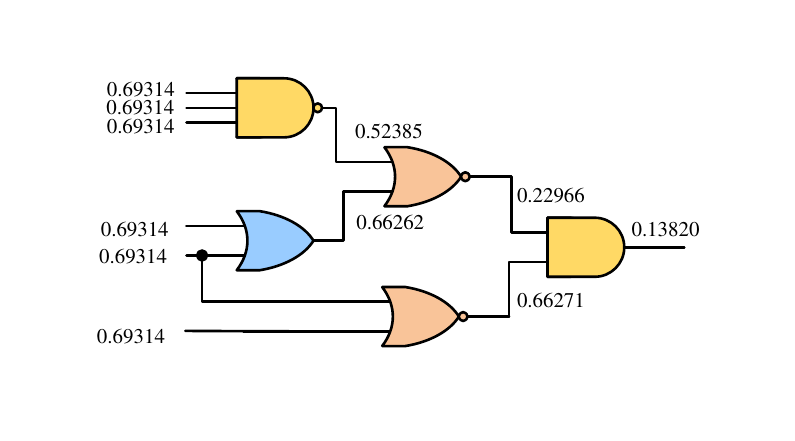}}
\caption{Information entropy of each wire in the given circuit fragment.}
\label{fig}
\end{figure}

\subsection{HT Detection based Clustering}
It's worth noting that our circuit analysis focuses on the state of internal wires in CUD rather than circuit structure. For the sake of the convenience of discussion, we define CUD = $<PI, W, POUT>$, where PI is the set of primary inputs, W is the set of internal signal wires and POUT is the set of primary outpus. More formally, PI = $\left\{pi_1, pi_2, ..., pi_l\right\}$, and W = $\left\{w_1, w_2, ..., w_m\right\}$ and POUT = $\left\{pout_1, pout_2, ..., pout_n\right\}$. After functional testing, we encode each original waveform of CUD and calculate the information entropy of each encoded waveform. Once the above step is complete, we apply a typical density-based clustering algorithm called Density-Based Spatial Clustering of Applications with Noise (DBSCAN) to perform HT detection in the information entropy space composed by W and POUT.

In the given data space, the density is defined as the number of data points within a specified radius (\textit{r}), and the \textit{core point} that has more than specified number of data points (\textit{MinPts}) within its r-neighborhood, and the \textit{border point} that has fewer than \textit{MinPts} within its r-neighborhood but it is in the r-neighborhood of a \textit{core point}, and and any point that is not a \textit{core point} or \textit{border point} is called \textit{noise point}. Moreover, date point q is \textit{directly density-reachable} from another point p, if p is a \textit{core point} and q is within the r-neighborhood of p. Data point q is \textit{density-reachable} from another point p, if there is a path of points $p_1$(p) $\to$ $p_2$ $\to$...$\to$ $p_{n-1}$ $\to$ $p_n$(q) such that point $p_{i+1}$ is \textit{directly density-reachable} from point $p_i$. Data point p and data point q are \textit{density-connected} if there is a data point o such that both p and q are \textit{density-reachable} from o.
\begin{algorithm}
\caption{HT detection based clustering}%算法名字
    \begin{algorithmic}[1]
        \Require Information entropy space, \textit{r}, \textit{MinPts}  % 算法的输入
        \Ensure Suspicious Trojan logics 
        \Repeat
        \State Select an unvisited data point (P) from information entropy space.
        \If {P is \textit{core point}}
        \State mark P as visited data point, then find all points which are \textit{density-reachable} from P, and form a cluster.
        \EndIf
        \If {P is \textit{border point}}
        \State mark P as visited data point, \Return{2}
        \EndIf
        \If {P is \textit{noise point}}
        \State delete P from information entropy space, \Return{2}
        \EndIf
        \Until{all data points in information entropy space have been visited}
        \State \textbf{Report} cluster with very low information entropy as suspicious Trojan logics.  % 返回结果
    \end{algorithmic}
\end{algorithm}

The basic idea of DBSCAN is to find the maximal sets of \textit{density-connected} points. That is, all points within the cluster are mutually \textit{density-connected}. Algorithm 1 shows the clustering process in the information entropy space.

\section{Test Pattern Generation for Suspicious Trojan Logics using Mutual Information}
In section 4, the proposed HT detection method can find suspicious Trojan logics. This section introduces a heuristic test pattern generation method using mutual information, which can further increase the transitions of suspicious Trojan logics. As is depicted in Figure 6, the correlation between each suspicious Trojan logic and each primary input is measured by mutual information. If the mutual information is greater than the threshold, corresponding primary input is referred to as strongly correlated primary input (SCPI) to this suspicious Trojan logic. Therefore, each suspicious Trojan logic will maintain a set of SCPI (SSCPI). Then, a heuristic algorithm is developed to select minimum SCPIs but to cover all suspicious Trojan logics.
\begin{figure}[htbp]
\centerline{\includegraphics[width=6.5cm,height=7.5cm]{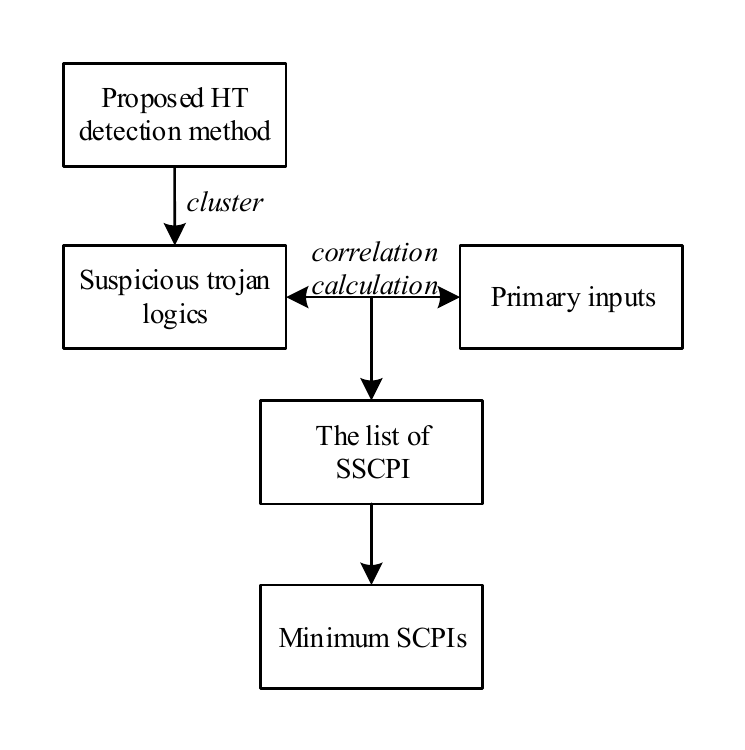}}
\caption{Overview of test patterns generation method.}
\label{fig}
\end{figure}

\subsection{Feasibility Analysis}
In the field of information theory, the mutual information between X and Y can measure the mutual dependence between the two variables. That is, mutual information can measure the correlation between two variables \cite{r33}. If X and Y are independent, their mutual information is zero. If X is a deterministic function of Y (Y also is a deterministic function of X), so knowing the value of X can determine the value of Y and vice versa. In this case, the mutual information between X and Y is the same as the H(X) and as the H(Y). 

Natively, each circuit logic can be expressed as a Boolean function of different primary inputs, which conforms statement of the correlation. For example, we can obtain three Boolean formula d = ab, e = $\overline{c}$ and f = ab + $\overline{c}$ for the given circuit structure, as shown in Figure 7. Hence, we can know that d and c, e and a, e and b, are independent such that their mutual information must be zero, and e is a deterministic function of c such that their mutual information is the same as H(c) and H(e), and the mutual information I(d; a) should be equal to the mutual information I(d; b) because of same circuit logic. It is worth noting that the mutual information I(f; a) is different from the mutual information I(f; c) because of different circuit logic (AND gate and Inverter). In short, the mutual information of two variables is higher, the correlation of variables is stronger.
\begin{figure}[htbp]
\centerline{\includegraphics[width=8cm,height=5cm]{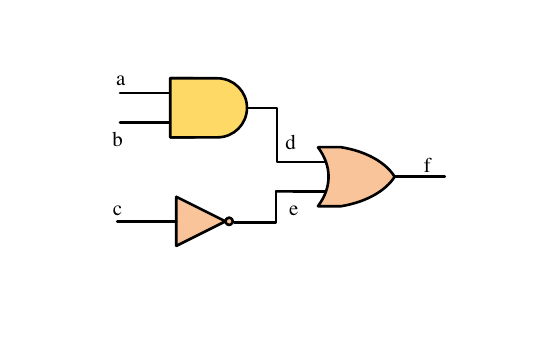}}
\caption{The example of mutual information analysis for circuit logic.}
\label{fig}
\end{figure}

\subsection{Correlation Calculation using Mutual Information}
We consider that the set of primary inputs PI = $\left\{pi_1, pi_2, ..., pi_l\right\}$, and consider that the set of suspicious Trojan logics (wires) SW = $\left\{sw_1, sw_2, ..., sw_t\right\}$, where t $\leq$ m+n. Firstly, we calculate mutual information I($sw_i$; $pi_j$) between each suspicious Trojan logic $sw_i$ and each primary input $pi_j$, where i = 1, 2, ..., t and j = 1, 2, ..., l. According to formula (7),  I($sw_i$; $pi_j$) = H($sw_i$) + H($pi_j$) - H($pi_j$, $sw_i$). Because each encoded waveform only contains 0 (non-transition) and 1 (transition),
H($pi_j$,$sw_i$) = - $\sum_{pi_j \in \left\{0,1\right\}}\sum_{sw_i \in \left\{0,1\right\}}p(pi_j,sw_i)\log_b p(pi_j,sw_i)$ 
according to formula (3). If I($sw_i$; $pi_j$) is greater than the threshold, we refer to the primary input $pi_j$ as the SCPI of suspicious Trojan logic $sw_i$. For each $sw_i$, its threshold is equal to $\sum_{pi_j \in PI} \frac{I(sw_i; pi_j)}{l}$, where l is the number of primary inputs. Finally, each suspicious trojan logic will have a SSCPI, and the strong correlation between primary inputs and suspicious trojan logics can constitute a strong correlation list as shown in Table 2.
\begin{table}[htbp]
\caption{The strong correlation list: 1 indicates $pi_j$ is a SCPI of $sw_i$ and 0 indicates no }
\begin{center}
\begin{tabular}{|c|c|c|c|c|c|}
\hline
 & $pi_1$ & $pi_2$ & $pi_3$ & ... & $pi_l$ \\
\hline
$sw_1$ & 1 & 0 & 1 & ... & 1 \\
\hline
$sw_2$ & 0 & 1 & 1 &... & 1 \\
\hline
... & ... & ... & ... & ... & ... \\
\hline
$sw_t$ & 1 & 1 & 0 & ... & 1 \\
\hline
\end{tabular}
\end{center}
\end{table}

\subsection{Test Patterns Generation}
Our goal is to select minimum number of SCPIs but to cover all suspicious Trojan logics. We define that $\left\{pi_j\right\}$ is set of suspicious Trojan logics whose SSCPI includes $pi_j$, and define `+' operation between sets is equivalent to the `union' operation between sets, and define `-' operation between sets is equivalent to the `difference' operation between sets. For example, $\left\{pi_1\right\}$ = $\left\{sw_1, sw_t\right\}$, $\left\{pi_l\right\}$ = $\left\{sw_1, sw_2, sw_t\right\}$, $\left\{pi_1\right\}$ + $\left\{pi_l\right\}$ = $\left\{sw_1, sw_2, sw_t\right\}$, and $\left\{pi_l\right\}$ - $\left\{pi_1\right\}$ = $\left\{sw_2\right\}$. Therefore, the problem can be abstracted as the following formula, where $x_j$ $\in$ $\left\{0, 1\right\}$. If $pi_j$ is selected, $x_j$ = 1, otherwise 0.

\begin{equation}
\begin{aligned}
    \min \sum_{pi_j \in PI} x_j \\
    \operatorname{ s.t. } \sum_{pi_j \in PI} x_j*\left\{pi_j\right\} = SW
\end{aligned}
\end{equation}

We develop a heuristic method to solve this problem. We define $f(k,y)$ indicates the optimal solution when $PI =  \left\{pi_1, ..., pi_k\right\}$ and $SW = y$. As shown in formula (9), it can be seen that $f(l,SW)$ is the optimal solution of formula (8). Algorithm 2 shows the core of solution. Then we perform constrained-random simulation, setting all the primary input at logic 0 or logic 1, which is not in SCPIs. For the rest of the primary inputs in SCPIs, we still generate full-random stimuli to perform simulation.

\begin{algorithm}
\small
    \caption{SCPI selection}%算法名字
    \begin{algorithmic}[1]
    \Require Strong correlation list  % 算法的输入
    \Ensure SCPIs       % 算法的输出
    \Function {f}{$l,SW$}
    \If {$pi_l \subseteq SW$}
        \State $ F(l,SW) = min\left\{ F(l-1,SW), F(l-1,SW-pi_l)+1 \right\} $
    \Else
        \State $F(l,SW) = F(l-1,SW)$
    \EndIf
    \EndFunction 
    \end{algorithmic}
\end{algorithm}

\begin{figure*}
\begin{equation}
\begin{aligned}
f(k,y) =
    \begin{cases} 
    \min \left\{ f(k-1,y), f(k-1, y-\left\{pi_k\right\})+1 \right\} & \mbox{if} \left\{pi_k\right\} \subseteq y \\
    f(k-1,y) & \mbox{otherwise } 
    \end{cases}
\end{aligned}
\end{equation}
\end{figure*}

\section{Experimens and Evaluations}
Proposed approach is evaluated on the different digital circuit designs from TrustHub benchmark \cite{r34}. All circuits are synthesized by Synopsys Design Compiler (DC) with Semiconductor Manufacturing International Corporation cell library for 90-nm silicon-on-insulator process. All circuits are simulated by Verilog Compiled Simulator (VCS) with high coverage as much as possible. We conduct data processing experiments and data analysis experiments on a computer with 2.8 GHz Intel Core i7 CPU and 8GB memory \cite{r35}. Brief information about the benchmarks used in our experiments is provided in Table 3.
\begin{table*}[htbp]
\caption{Brief informations of circuits under detection}
\begin{center}
\begin{tabular}{|c|c|c|}
\hline
 Circuit & \# units & Features of HT \\
\hline
RS232\_T1000 & 215 & Trojan trigger is a combinational comparator; change functionality\\
\hline
RS232\_T1100 & 217 &  Trojan trigger is a sequential comparator; change functionality\\
\hline
RS232\_T1200 & 216 &  Trojan trigger is a sequential comparator; change functionality\\
\hline
RS232\_T1300 & 213 &  Trojan trigger is a combinational comparator; change functionality\\
\hline
RS232\_T1400 & 215 &  Trojan trigger is a sequential comparator; change functionality\\
\hline
RS232\_T1500 & 216 &  Trojan trigger is a sequential comparator; change functionality\\
\hline
RS232\_T1600 & 214 &  Trojan trigger is a sequential comparator; change functionality\\
\hline
s15850\_T100 & 2182 & Trojan trigger consists of two comparators and two flip-flops; leak an internal signal. \\
\hline
s35932\_T200 & 5438 & Trojan trigger is a comparator; denial of Service. \\
\hline
s38417\_T100 & 5341 &  Trojan trigger is a comparator; change functionality, denial of service. \\
\hline
\end{tabular}
\end{center}
\end{table*}

\subsection{Clustering Comparison between Information Entropy Space and Transition Probability Space}
In our experiments, our method can detect all suspicious Trojan logics in the CUD. Taking RS232\_T1000 and RS232\_T1100 as examples, we present the difference of clustering between information entropy space and transition probability space.
Figure 8(a) and Figure 8(b) shows the result of clustering for RS232\_T1000 benchmark and RS232\_T1100 benchmark, respectively. It is worth noting that the clustering process only focuses on the \textit{density-reachable} relationship of information entropy space.

\begin{figure}[htbp]
\centering
\subfigure[Clustering for RS232\_T1000 benchmark.]{
\includegraphics[width=8cm,height=4cm]{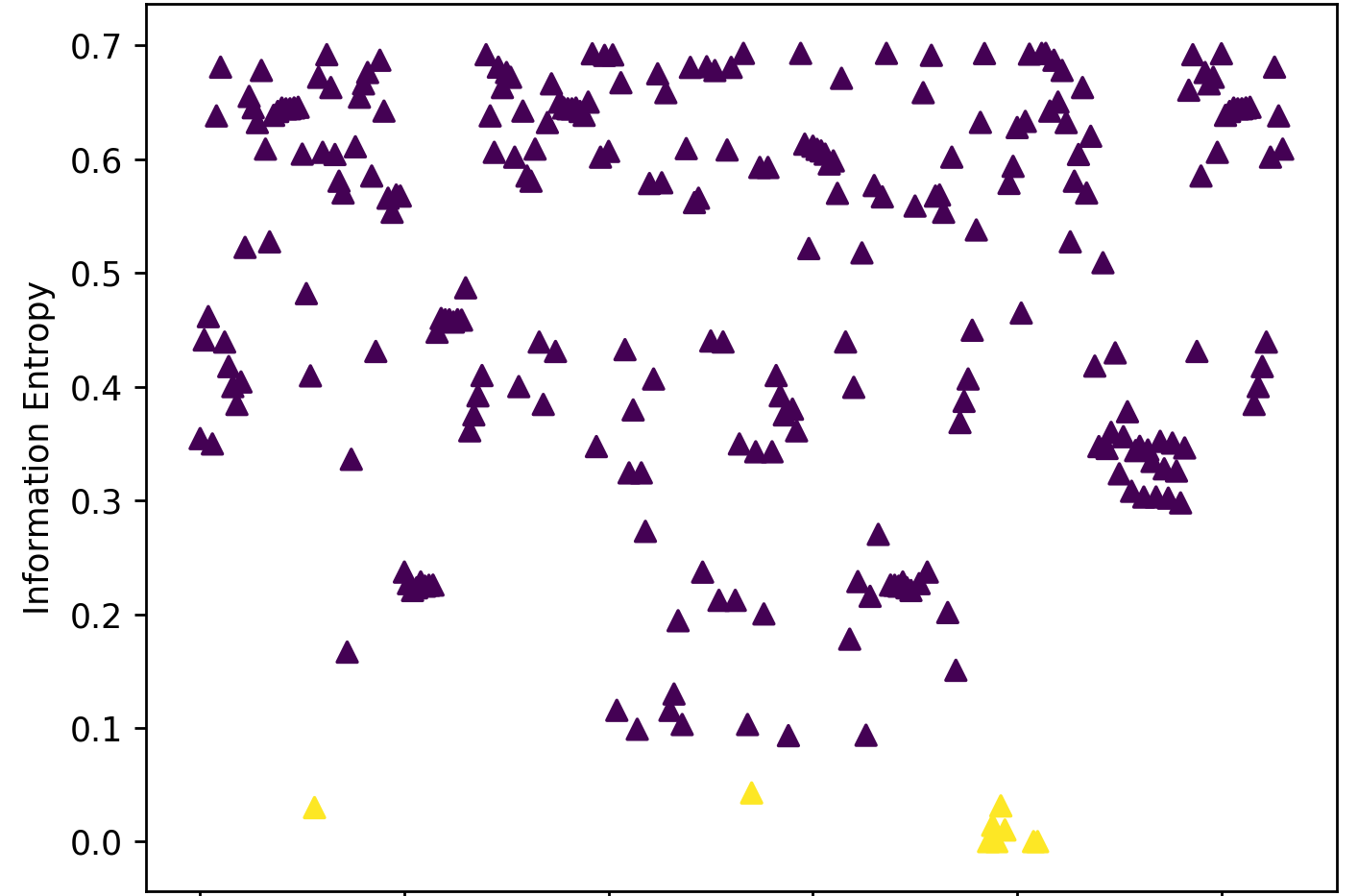}
}
\subfigure[Clustering for RS232\_T1100 benchmark.]{
\includegraphics[width=8cm,height=4cm]{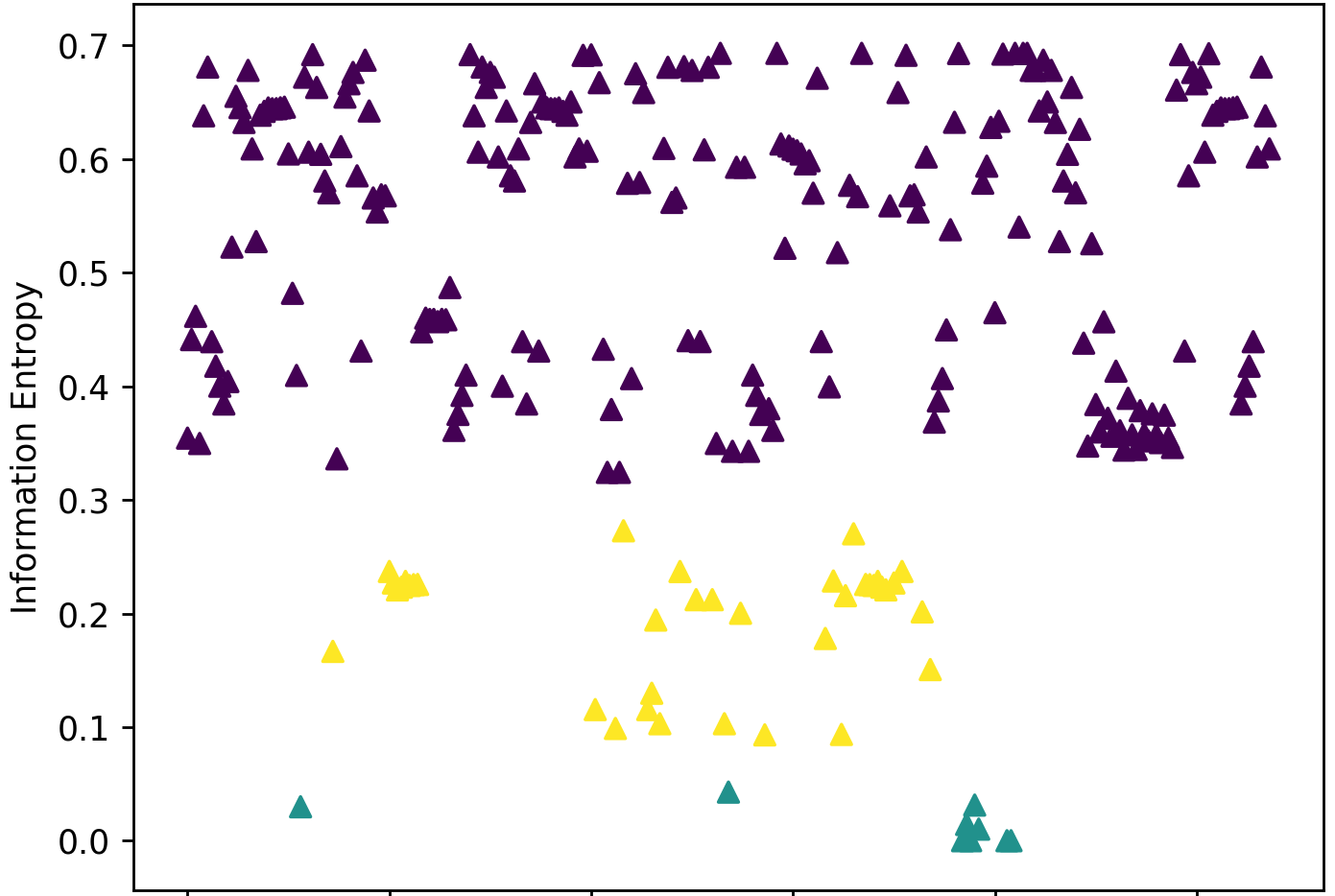}
}
\caption{Clustering in information entropy space.}
\end{figure}

As shown in Figure 8, though the clustering algorithm can divide the information entropy space into several cluters (2 or 3), the circuit logics with extremely low information entropy are always divided into one cluster according to the \textit{density-reachable} relationship. Similarly, we also use transition probability for Trojan detection. Under the same parameters, Figure 9(a) and Figure 9(b) shows the result of clustering for RS232\_T1000 and RS232\_T1100, respectively.

\begin{figure}[htbp]
\centering
\subfigure[Clustering for RS232\_T1000 benchmark.]{
\includegraphics[width=8cm,height=4.2cm]{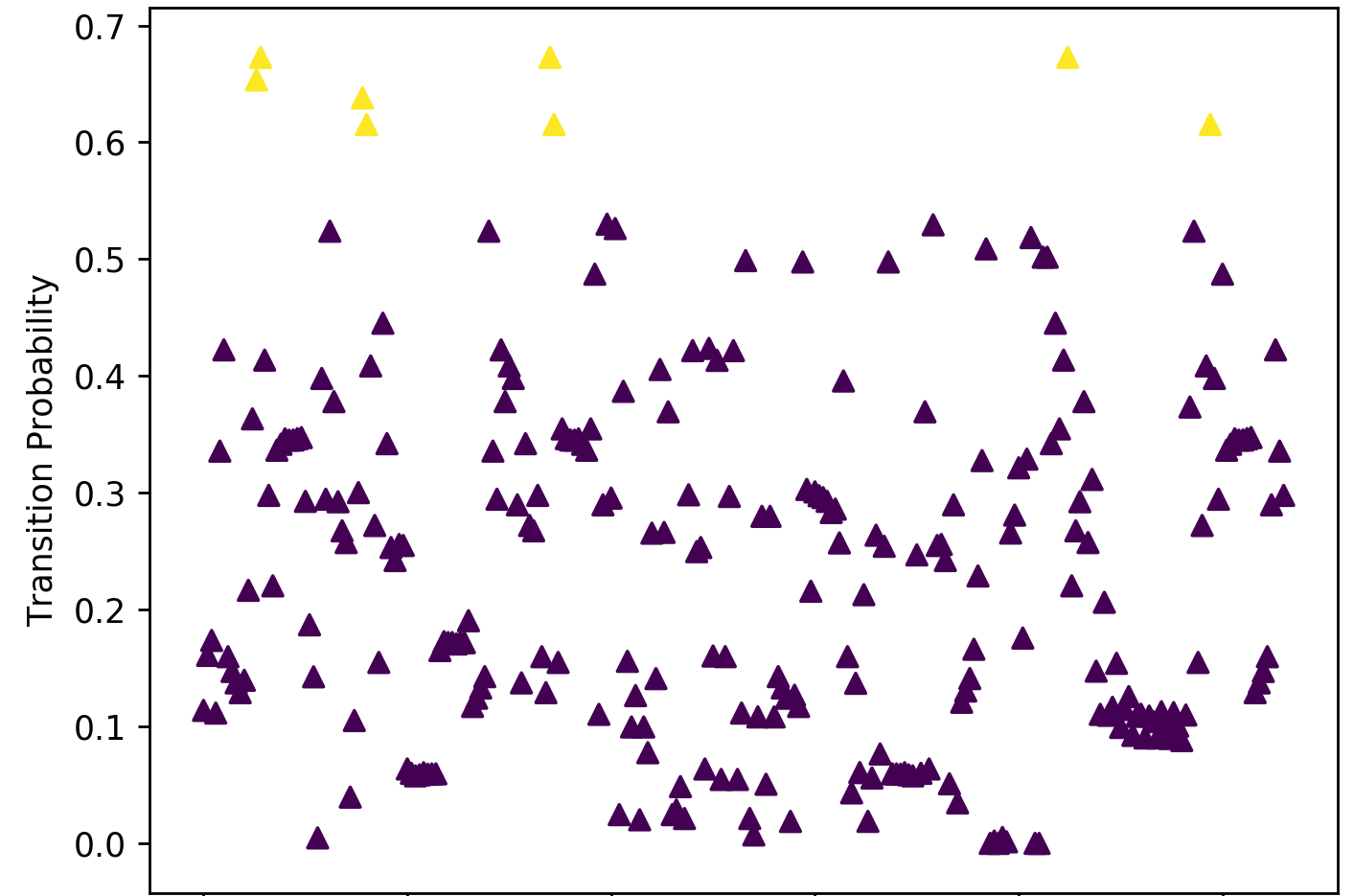}
}
\subfigure[Clustering for RS232\_T1100 benchmark.]{
\includegraphics[width=8cm,height=4.2cm]{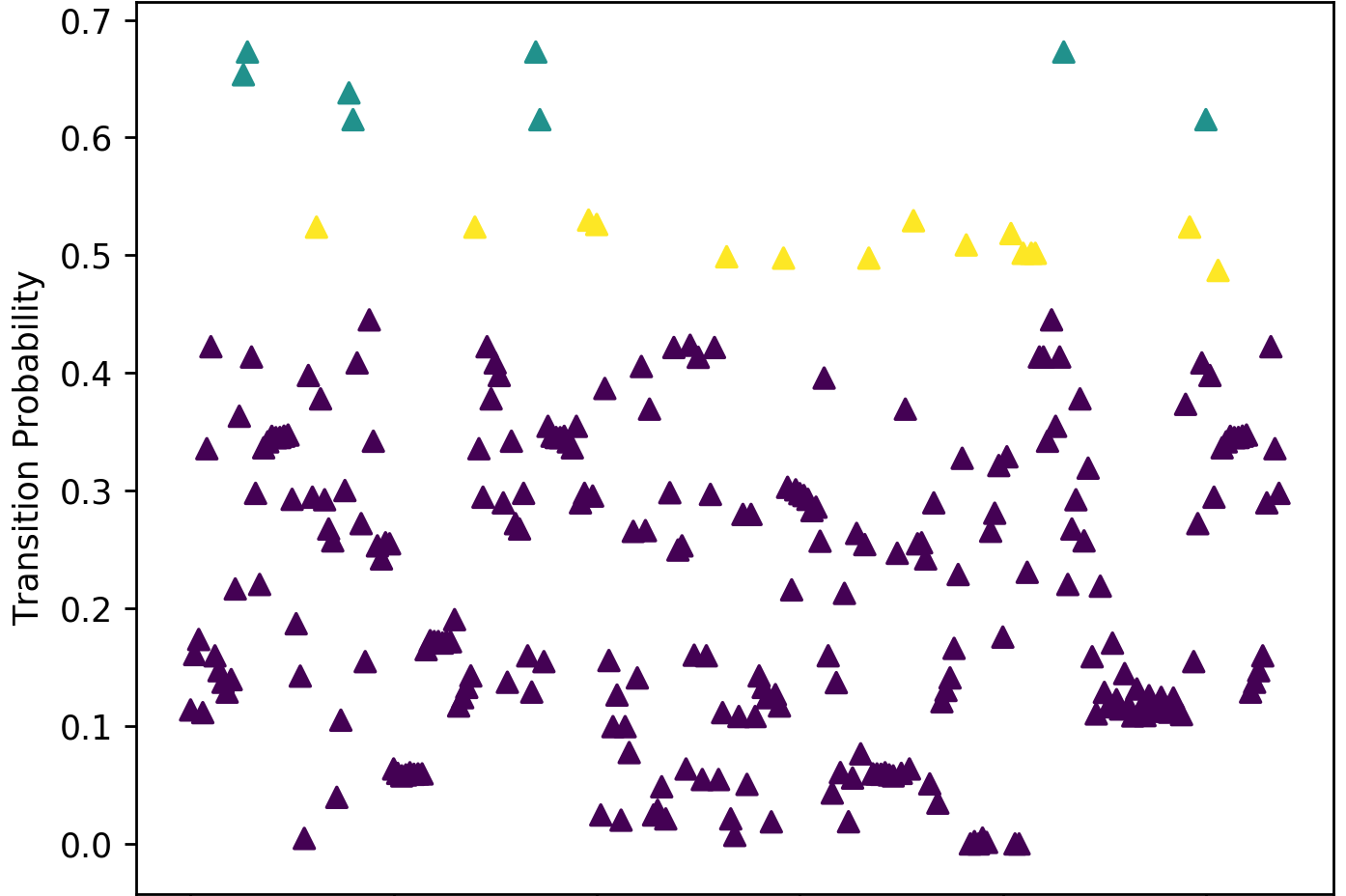}
}
\caption{Clustering in transition probability space.}
\end{figure}

It can be seen that transitions will result in high false positives. However, the information entropy can effectively distinguish the difference between Trojan logics and normal logics. In order to have a more intuitive insight on the difference between information entropy and transition probability, we sort the information entropy space and transition probability space of RS232\_T1000 benchmark from lowest to highest, respectively. Then the distribution of information entropy and transition probability are shown in Figure 10. As shown in Figure 10(a), the area with low information entropy (red) and other area (green) have obvious \textit{density-unreachable} relationship. However, the area with low transition probability and other area are still \textit{density-reachable} (red) in transition probability space shown in Figure 10(b), which will lead to poor Trojan detection performance. Because the information entropy can amplify the difference between low transition probability and high transition probability, it can detect effectively suspicious Trojan logics.
\begin{figure}[htbp]
\centering
\subfigure[Sorted distribution of information entropy.]{
\includegraphics[width=8cm,height=4.2cm]{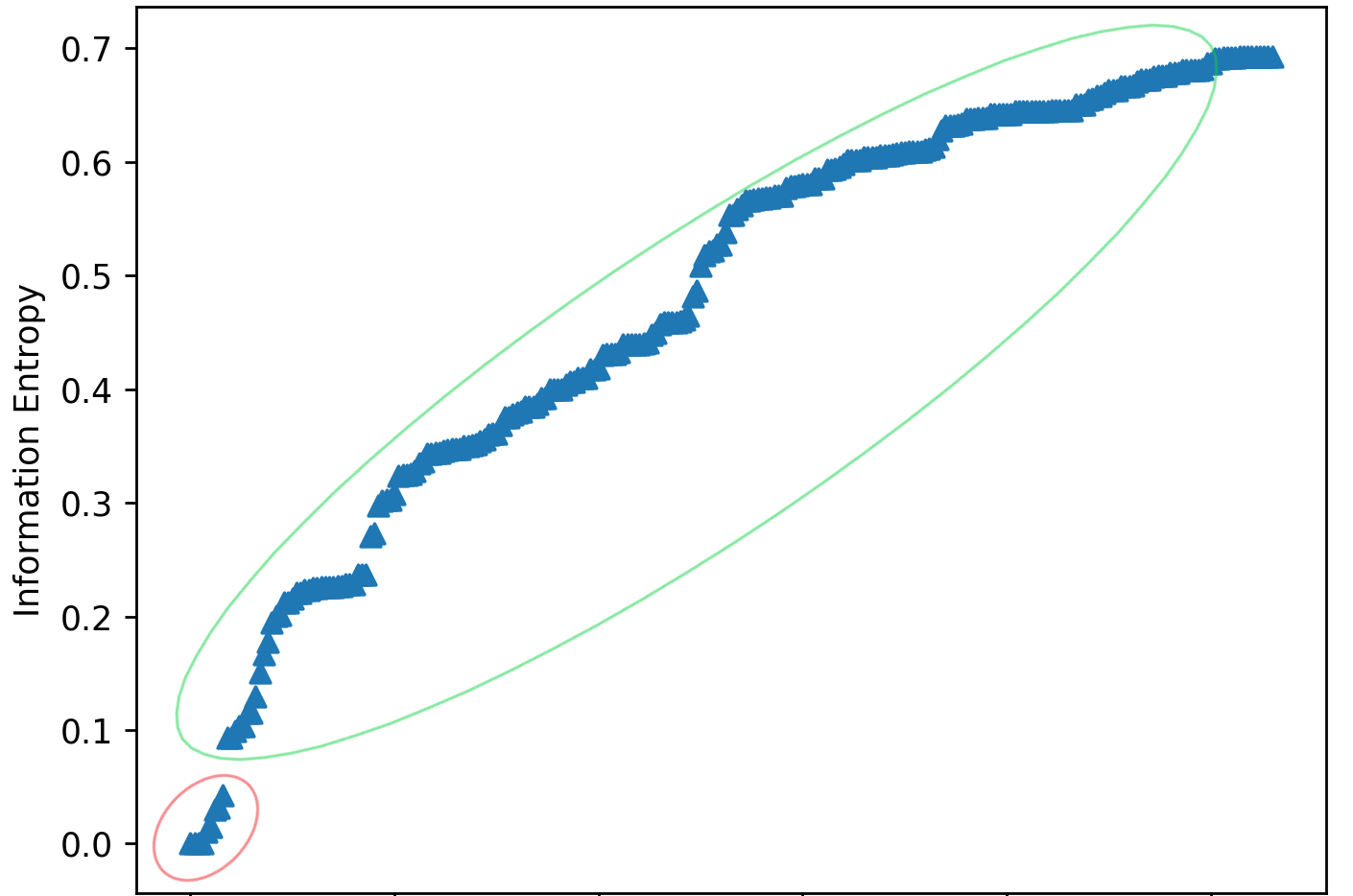}
}
\subfigure[Sorted distribution of transition probability.]{
\includegraphics[width=8cm,height=4.2cm]{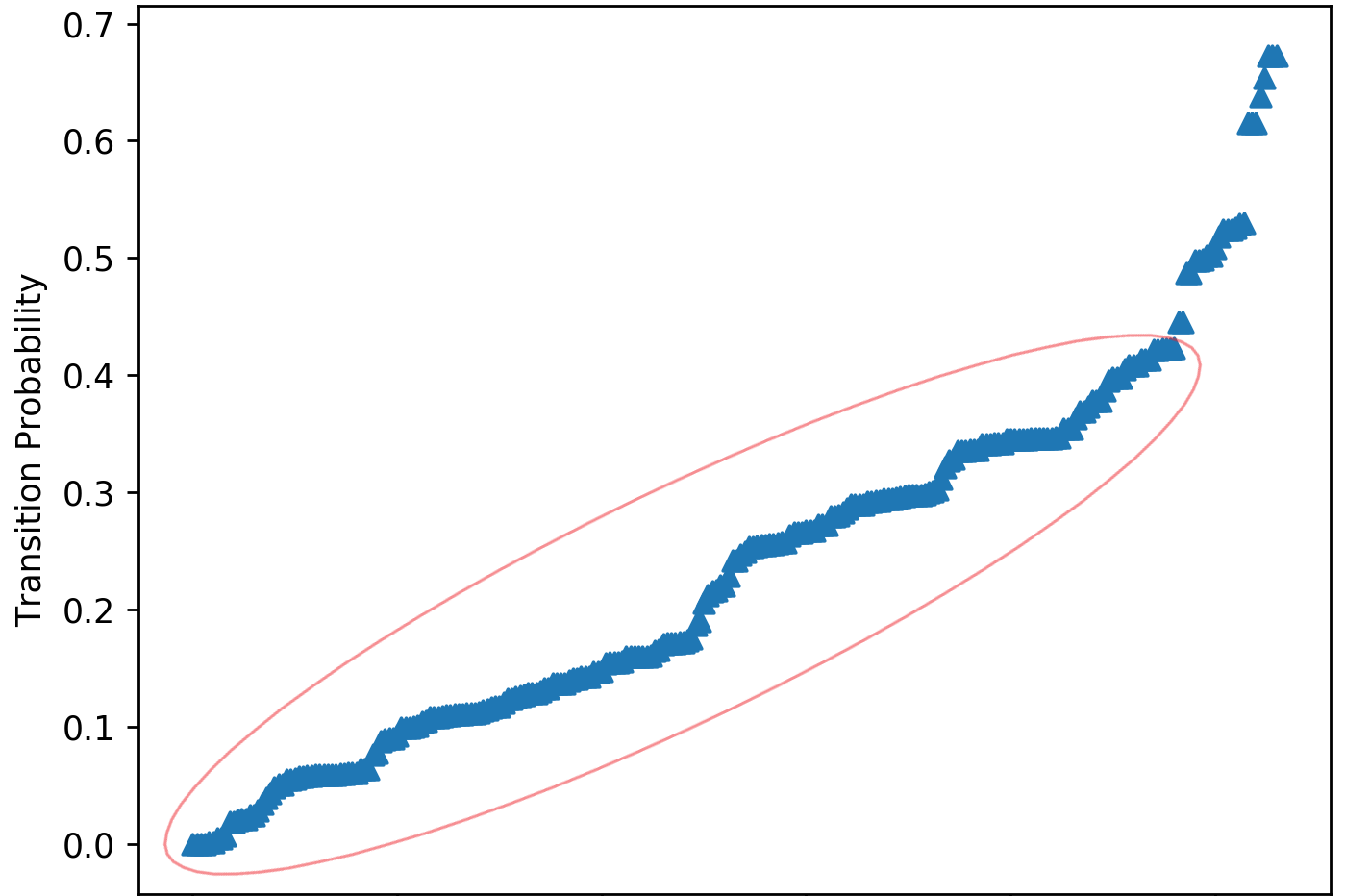}
}
\caption{Difference between information entropy space and transition probability space for RS232\_T1000 benchmark.}
\end{figure}

\subsection{HT Detection Performance and Parameter Analysis}

To further evaluate the effectiveness of HTDet, we manually check the suspicious Trojan logics reported by the clustering algorithm. The results are shown in Table 4. \textit{MinPts} and \textit{r} are the parameters used in clustering process.
The sensitivity of the results is measured by the true positive rate (TPR), i.e. the number of Trojan wires correctly detected as a percentage of the total number of Trojan logics. We also provide the true negative rage (TNR) results, which tells us the ratio of the true negatives over the number of non-Trojan logics. False positive rate (FPR = 1 - TNR) is the fraction of logics that are falsely flagged as being suspicious Trojan logics. It can be seen that HTDet can effectively detect Trojan logics of CUD with the extremely low false positives.

We also analyze the effect of parameters \textit{MinPts} and \textit{r} on HT detection performance using control variable method. When \textit{r} is fixed to 0.05, both TPR and TNR decline as \textit{MinPts} increases, as shown in Figure 11(a). This is because the number of \textit{noise point} gradually increases when \textit{MinPts} increases. Similarly, when \textit{MinPts} is fixed to 5 and \textit{r} increases, TPR gradually decline but TNR almost is constant, as shown in Figure 11(b). This is because all data points are clustered into normal logcis when r is equal to 0.06 or 0.07. Hence, the appropriate values of parameters are also necessary for Trojan detection.

\begin{table}[htbp]
\caption{Results of manual check}
\begin{center}
\begin{tabular}{|c|c|c|c|c|c|c|}
\hline
 Circuit & \textit{MinPts} & \textit{r} & TPR & TNR \\
\hline
RS232\_T1000 & 2 & 0.05 & 62\% & 99\% \\
\hline
RS232\_T1100 & 5 & 0.04 & 67\% & 99\% \\
\hline
RS232\_T1200 & 5 & 0.04 & 89\% & 99\% \\
\hline
RS232\_T1300 & 2 & 0.05 & 89\% & 99\% \\
\hline
RS232\_T1400 & 5 & 0.04 & 61\% & 99\% \\
\hline
RS232\_T1500 & 5 & 0.04 & 73\% & 99\% \\
\hline
RS232\_T1600 & 5 & 0.04 & 62\% & 99\% \\
\hline
s15850\_T100 & 4 & 0.05 & 96\% & 99\% \\
\hline
s35932\_T200 & 5 & 0.05 & 93\% & 99\% \\
\hline
s38417\_T100 & 4 & 0.05 & 100\% & 99\% \\
\hline
\end{tabular}
\end{center}
\end{table}

\begin{figure}[htbp]
\centering
\subfigure[The effect of \textit{MinPts} on TPR and TNR.]{
\includegraphics[width=7cm,height=4.5cm]{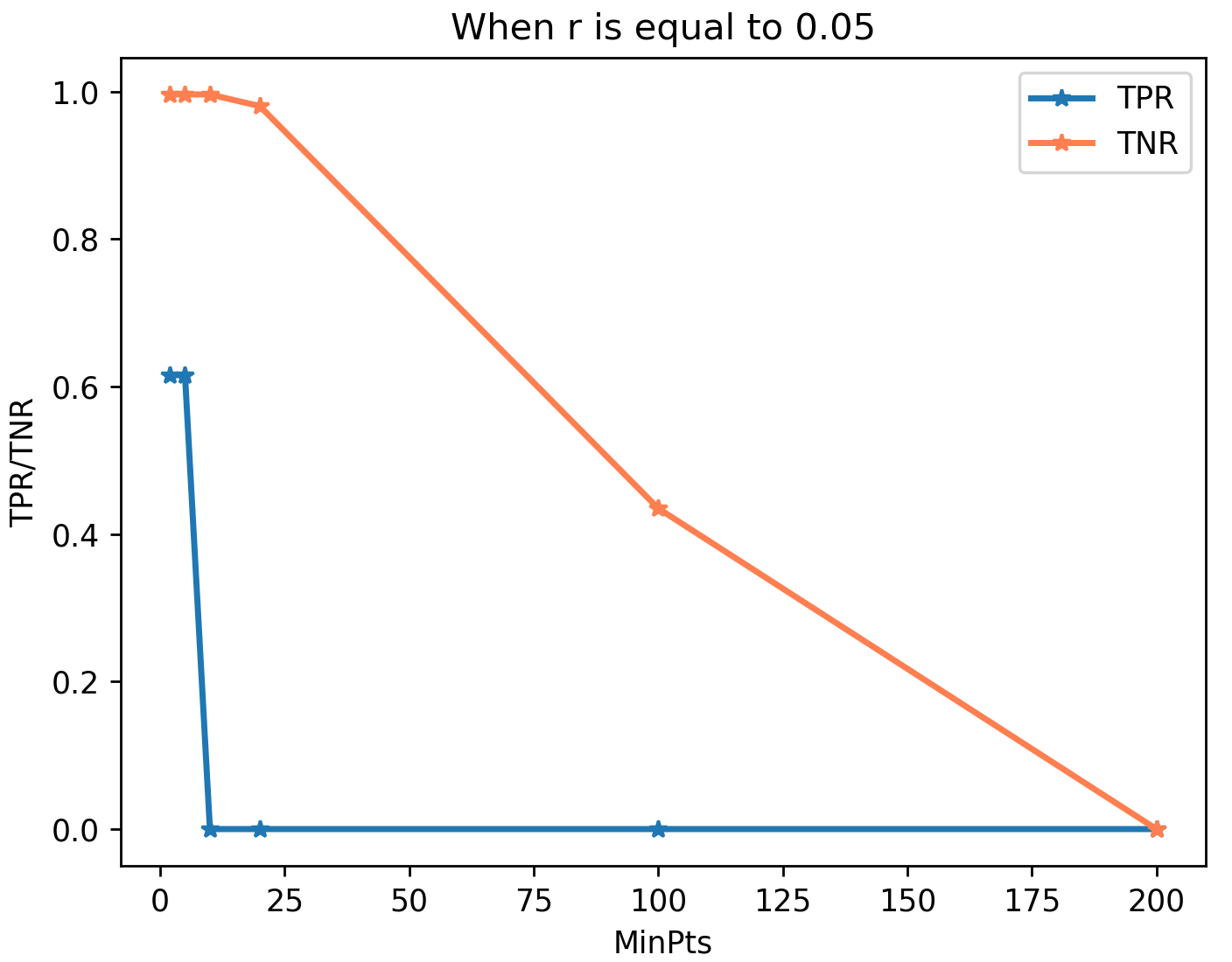}
}
\subfigure[The effect of \textit{r} on TPR and TNR.]{
\includegraphics[width=7cm,height=4.5cm]{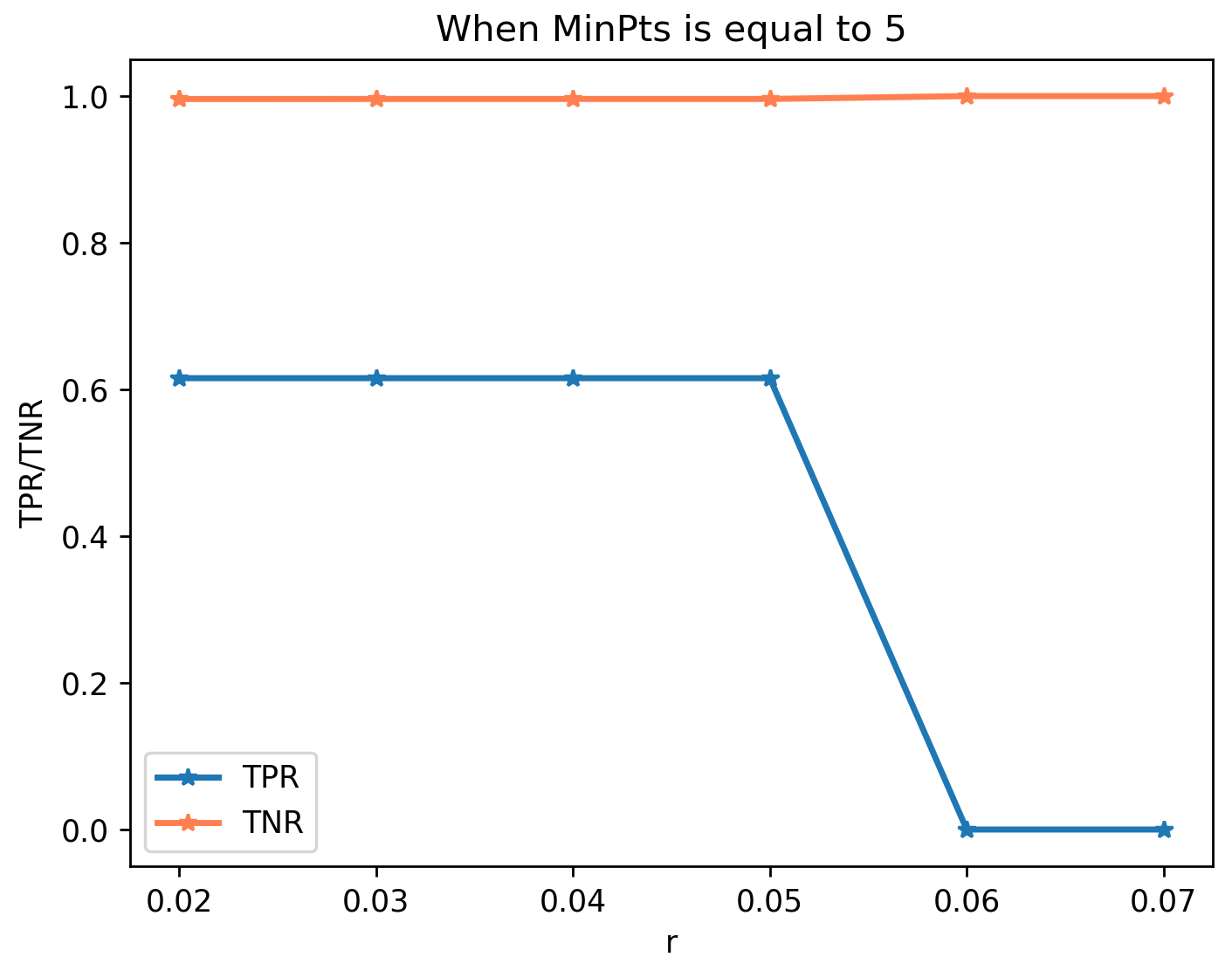}
}
\caption{Parameter Analysis on RS232\_T1000 benchmark.}
\end{figure}

\subsection{Comparison to existing methods}

we compare the experimental results to existing methods in the point of TPR and TNR. Reference \cite{r16} proposed a HT detection method based on static structure analysis, and Reference \cite{r23} proposed a HT detection method based on signal correlations. Table 5 shows the comparison to [16], and Table 6 shows the comparison to [23].
Obviously, our approach can obtain better HT detection performance in order to achieve the good trade-off between TPR and TNR. In the point of average TNR, it can obtain the 99\% average TNR value, which indicates that proposed technique, HTDet, can significantly reduce false positives.

\begin{table}[htbp]
\caption{Comparison to the existing method \cite{r16}}
\begin{center}
\begin{tabular}{|c|c|c|c|c|}
\hline
  & \multicolumn{2}{|c|}{TPR} & \multicolumn{2}{|c|}{TNR} \\
\hline
Circuit & [16] & Ours & [16] & Ours \\
\hline
RS232\_T1000 & 53\% & 62\% & 31\% & 99\% \\
\hline
RS232\_T1100 & 58\% & 67\% & 27\% & 99\% \\
\hline
RS232\_T1200 & 80\% & 89\% & 26\% & 99\% \\
\hline
RS232\_T1300 & 89\% & 89\% & 26\% & 99\% \\
\hline
RS232\_T1400 & 83\% & 61\% & 22\% & 99\% \\
\hline
RS232\_T1500 & 83\% & 73\% & 24\% & 99\% \\
\hline
RS232\_T1600 & 89\% & 62\% & 26\% & 99\% \\
\hline
s15850\_T100 & 93\% & 96\% & 66\% & 99\% \\
\hline
s35932\_T200 & 100\% & 93\% & 59\% & 99\% \\
\hline
s38417\_T100 & 100\% & 100\% & 76\% & 99\% \\
\hline
Average & \textbf{83\%} & 79\% & 39\% & \textbf{99\%} \\ 
\hline
\end{tabular}
\end{center}
\end{table}

\begin{table}[htbp]
\caption{Comparison to the existing method \cite{r23} }
\begin{center}
\begin{tabular}{|c|c|c|c|c|}
\hline
  & \multicolumn{2}{|c|}{TPR} & \multicolumn{2}{|c|}{TNR} \\
\hline
Circuit & [23] & Ours & [23] & Ours \\
\hline
s15850\_T100 & 61\% & 96\% & 99\% & 99\% \\
\hline
s35932\_T200 & 27\% & 93\% & 99\% & 99\% \\
\hline
s38417\_T100 & 100\% & 100\% & 99\% & 99\% \\
\hline
Average & 63\% & \textbf{96\%} & 99\% & 99\% \\ 
\hline
\end{tabular}
\end{center}
\end{table}
In this paper, we do not attempt to find all Trojan logics (wires), but try the best to find a set of most suspicious logics, which can effectively reduce the authentication time. That is, a manual check after the automatic HT detection is always necessary.

\subsection{Effectiveness Analysis of Test Patterns Generation Method}
We randomly selected three benchmarks (RS232\_T1000, RS232\_T1100 and s15850\_T100) to evaluate the effectiveness of proposed test patterns generation method. Let the transition of each suspicious logic $sw_i$ be $tr_i$ during the simulation, where $sw_i \in SW$, and i = 1, 2, ..., t. Let $tr_{max}$ be the maximum of $tr_i$. Let $tr_{ave}$ be equal to $\frac{\sum_{i=1}^t tr_i }{t}$. Then, maximum transition and average transition are used to measure the effectiveness of test patterns.
After obtaining SCPIs, we set that all the primary inputs, which are not in SCPIs, at logic 0 or logic 1. For the primary inputs in SCPIs, we still generate full-random stimuli to perform simulation. After $10^6$ cycles of simulation, the transitions of suspicious Trojan logics are summarized in Table 7.

It can be seen that proposed test patterns generation method can increase effectively the maximum transition and average transition of these suspicious logics, which means that it can reduce activation time.
\begin{table}[htbp]
\caption{Transitions Comparison: Before\_* indicates full-random test stimuli, and After\_* indicates constrained-random test stimuli using our approach }
\begin{center}
\begin{tabular}{|c|c|c|}
\hline
 Circuit & $tr_{max}$ & $tr_{ave}$\\
\hline
Before\_RS232\_T1000 & 722 & 224.67\\
\hline
After\_RS232\_T1000 & 768 & 230.89\\
\hline
Before\_RS232\_T1100 & 719 & 224.39 \\
\hline
After\_RS232\_T1100 & 746 & 231.56\\
\hline
Before\_s15850\_T100 & 716 & 64.19\\
\hline
After\_s15850\_T100 & 954 & 96.48\\
\hline
\end{tabular}
\end{center}
\end{table}

\section{Related Works}
HT detection is a challenging problem. Lots of researches on HT detection have been proposed in the past decades, which can be roughly classified into reverse engineering, side channel analysis, static structure analysis, statistical feature analysis and functional testing.

Bao proposed that using reverse engineering to dissect the chip under detection can guarantee that any malicious modifications in chip can be detected \cite{r5,r6}. However, the limitation of this method is that the time cost is too much, it even takes several weeks to analyze the chip under detection. Hence, the reverse engineering can only be applied to the IC with small scale and simple structure. 

In side channel analysis \cite{r7,r8,r9,r10,r11,r12,r13}, the impacts of HTs (e.g., circuit delay, transient current, leakage power and heat analysis) are used to detect whether there are the HTs in CUD. However, the characteristics of circuit is more susceptible to process variations and environmental noise due to the present nanoscale technologies \cite{r14}.

 A score-based classification method is proposed for identifying HTs in CUD \cite{r15}. This technique comprehensively analyzes the characteristics of Trojan logics introduced at TrustHub \cite{r34}, then uses a strategy of conditional judgment for HT detection. Hasegawa proposed learning structure features for Trojan detection \cite{r16,r17,r18}. For this purpose, support vector machine, multi-layer neural network and random forest is applied to learn circuit structure features, respectively. Reference \cite{r19} summarized the triggering characteristics of Trojan circuits and proposed a feature analysis technique based on flip-flop level information flow graph. Then, a multilevel HT detection framework is proposed \cite{r20}, which combines flip-flop level and combinational logic level structure feature analysis.

 Reference \cite{r21} analyzes time to generate a transition in functional Trojans. Transition is modeled by geometric distribution and the number of clock cycles required to generate a transition is estimated. FANCI \cite{r22} considers that the input-to-output dependency has significant difference between Trojan logic and normal logic, so it flags logics which have weak input-to-output dependency as suspicious Trojan logics by Boolean function analysis. In \cite{r23}, a HT detection method using signal correlation has been proposed. It basically estimates the statistical correlation between signals in a circuit for Trojan detection with the use of ordering points to identify the clustering structure algorithm. Furthermore, \cite{r24} proposed a reference-free HT detection scheme based on controllability and observability. This paper indicates that the characteristics of controllability and observability between Trojan gates and genuine gates have significant difference. In \cite{r25}, a novel HT detection approach through distinguishing the ``unnaturalness" of HT from the ``naturalness" of normal circuits by applying natural language processing technology. This paper considers that design teams of commercial chips will have the specific design style due to the existence of established design specifications, so the statistical method can be used to detect abnormal circuit logics.

Functional testing-based HT detection approaches \cite{r26,r27,r28,r29} try to generate random test patterns to activate the HTs in CUD. If the logical values of primary outputs do not match the correct results, a Trojan is detected. The primary challenge of functional testing-based method is that the Trojan circuit is much smaller than the original circuit, and HTs usually have the dormant nature. Hence, it is difficult to detect potential HTs in CUD by traditional functional testing.

Different from traditional functional verification approaches, we propose HTDet, a novel HT detection technique based on information entropy. We consider that the Trojan usually be inserted in the regions with low controllability and low observability in order to maintain high concealment, which will result in that Trojan logics appear extremely low transitions during the simulation. Our approach does not require that the Trojan logic is pushed the triggering state. As long as the transitions of circuit logics are extremely low, HTDet can flag them as suspicious Trojan logics using \textit{density-reachable} relationship. Although the information theory has been applied in many fields, to the best of our knowledge, this is the first attempt to use the information theory technology to detect HTs in hardware design.

\section{Conclusions}
In this paper, we propose a novel HT detection method named HTDet, which can distinguish effectively the transitions difference betwwen normal logics and Trojan logics using information entropy technique. HTDet is an unsupervised learning method and can find quickly suspicious Trojan logics without the requirement on the ``Golden Circuit". HTDet does not require that the Trojan logic is pushed the activation state during the simulation, and it flags circuit logics with extremely low information entropy as suspicious Trojan logics. Besides, we develop a heuristic method to increase transitions of suspicious Trojan logics using mutual information. Experimental results demonstrate the effectiveness of HTDet.

%\bibliographystyle{IEEEtran}
%\bibliography{ref}

\end{document}